\documentclass[aps,twocolumn,superscriptaddress]{revtex4-2}

\usepackage{graphicx}  % needed for figures
\usepackage{float}     % needed for non-float figures
\usepackage{dcolumn}   % needed for some tables
\usepackage{bm}        % for math
\usepackage{amssymb}   % for math
\usepackage{amsmath}   % for math
\usepackage{lipsum}    % One column equation in two-column document class
\usepackage{xcolor}
\usepackage{soul}
\usepackage[utf8]{inputenc}
\usepackage{natbib}
\usepackage{graphicx}

\usepackage{color}

\definecolor{goodred}{RGB}{183,15,58}
\definecolor{goodblue}{RGB}{93,128,180}

\begin{document}
\title{Theory of inertial spin dynamics in anisotropic ferromagnets}
\author{Mikhail Cherkasskii}
\email[]{macherkasskii@hotmail.com}
\affiliation{Faculty of Physics and Center of Nanointegration (CENIDE), University of Duisburg-Essen, Duisburg 47057, Germany}

\author{Igor Barsukov}
\email[]{igorb@ucr.edu}
\affiliation{Department of Physics and Astronomy, University of California, Riverside, California 92521, USA}

\author{Ritwik Mondal}
\altaffiliation{Present address: Department of Physics, Indian Institute of Technology (ISM) Dhanbad, IN-826004, Dhanbad, India}
\affiliation{Department of Spintronics and Nanoelectronics, Institute of Physics ASCR, v.v.i., Cukrovarnická 10, Prague 6, 162 53, Czech Republic}
\affiliation{Department of Physics and Astronomy, Uppsala University, Box 516, SE-75120 Uppsala, Sweden}

\author{Michael Farle}
\affiliation{Faculty of Physics and Center of Nanointegration (CENIDE), University of Duisburg-Essen, Duisburg 47057, Germany}

\author{Anna Semisalova}
\email[]{anna.semisalova@uni-due.de}
\affiliation{Faculty of Physics and Center of Nanointegration (CENIDE), University of Duisburg-Essen, Duisburg 47057, Germany}

\begin{abstract}
Recent experimental observation of inertial spin dynamics calls upon holistic reevaluation of the theoretical framework of magnetic resonance in ferromagnets. Here, we derive the secular equation of an inertial spin system in analogy to the ubiquitous Smit-Beljers formalism. We find that the frequency of precessional ferromagnetic resonances is decreased as compared to the noninertial case.  We also find that the frequency of nutational resonances is generally increased due to the presence of magnetic anisotropy and applied magnetic field. We obtain exact solutions of the secular equation and approximations that employ the terminology of non-inertial theory and thus allow for convenient estimates of the inertial effects.
\end{abstract}
\maketitle

\section{Introduction}
Inertial effects of spin dynamics have mostly been neglected in the analysis of experimental spin resonance and spin transport results, since they mainly manifest themselves at hardly accessible terahertz frequencies in ferromagnets. With recent advances in spectroscopic techniques \cite{Fulop_Tzortzakis_Kampfrath_2020, Seifert_Jaiswal2016}, this fundamental phenomenon has been observed in permalloy, cobalt and cobalt-iron-boron ferromagnetic films \cite{Neeraj_Awari2021,Unikandanunni2021}. Inertial spin dynamics offers novel avenues for ultrahigh-frequency spintronic applications using well-established ferromagnetic materials and is thus becoming a prominent field of research.

The mathematical concept of inertia of magnetization was introduced in the context of magnetoelastic coupling in ferromagnets by Suhl \cite{Suhl_1998}. It was followed by an extension of the breathing Fermi surface model which demonstrated the emergence of a damping contribution linked to inertia \cite{Fahnle_Steiauf_Illg_2011, fahnle2013erratum}. It was noticed that the Landau-Lifshitz-Gilbert equation, which describes magnetization motion in analogy with a spinning top, required an inertial tensor of a rigid body. A revision \cite{Wegrowe_Ciornei_2012} of this analogy within a macroscopic Lagrangian approach suggested that inertia originates from generalization of gyromagnetic ratio –- the magnetic moment is non-collinear to the angular momentum. The Landau-Lifshitz-Gilbert equation was extended by including the second-order time derivative of magnetization, which now resembles Newton’s equation of motion for a massive point particle.

This inertial Landau-Lifshitz-Gilbert (ILLG) equation reads
	\begin{equation} \label{eq:ILLG}
	\begin{split}
	{\partial _t}{\bf{M}} &=  - \left| \gamma  \right|{\mu _0}{\bf{M}} \times {{\bf{H}}_{{\rm{eff}}}} + \dfrac{\alpha }{{{M_0}}}{\bf{M}} \times {\partial _t}{\bf{M}} \\
	&+ \dfrac{\eta }{{{M_0}}}{\bf{M}} \times {\partial _{tt}}{\bf{M}},
	\end{split}
	\end{equation}
where $\gamma  = g{\mu _{\rm{B}}}/\hbar$ is the gyromagnetic ratio, ${\mu _{\rm{B}}}$ is the Bohr magneton, $g$ is the g-factor, $\mu _0$ is the permeability of free space, ${\bf{M}}$ is the magnetization with the magnitude ${M_0}$, ${{\bf{H}}_{{\rm{eff}}}}$ is the effective magnetic field, $\alpha $ is the Gilbert damping, and $\eta $ is the inertial parameter. The inertial parameter has been discussed to be correlated to the Gilbert damping within the breathing Fermi surface model \cite{Fahnle_Steiauf_Illg_2011, fahnle2013erratum} and torque-torque correlation model \cite{Thonig_Eriksson_Pereiro_2017}, whereas inertia has been considered independent of damping within the classical Lagrangian approach \cite{Wegrowe_Ciornei_2012}. The ILLG equation was derived within the framework of mesoscopic nonequilibrium thermodynamics \cite{Ciornei_Rubi_Wegrowe_2011} and the Dirac-Kohn-Sham theory \cite{Mondal_Berritta_Oppeneer_2018}. In an atomistic method, exchange interaction, damping, and inertia were calculated from first principles \cite{Bhattacharjee_Nordstrom_Fransson_2012}. The microscopic origin of inertia has been asserted in the relativistic spin–orbit coupling \cite{Mondal_Berritta_Carva_Oppeneer_2015,Mondal_Berritta_Nandy_Oppeneer_2017,Thonig_Eriksson_Pereiro_2017}.

\begin{figure}[t]
    \centering
    \includegraphics[width=0.40\textwidth]{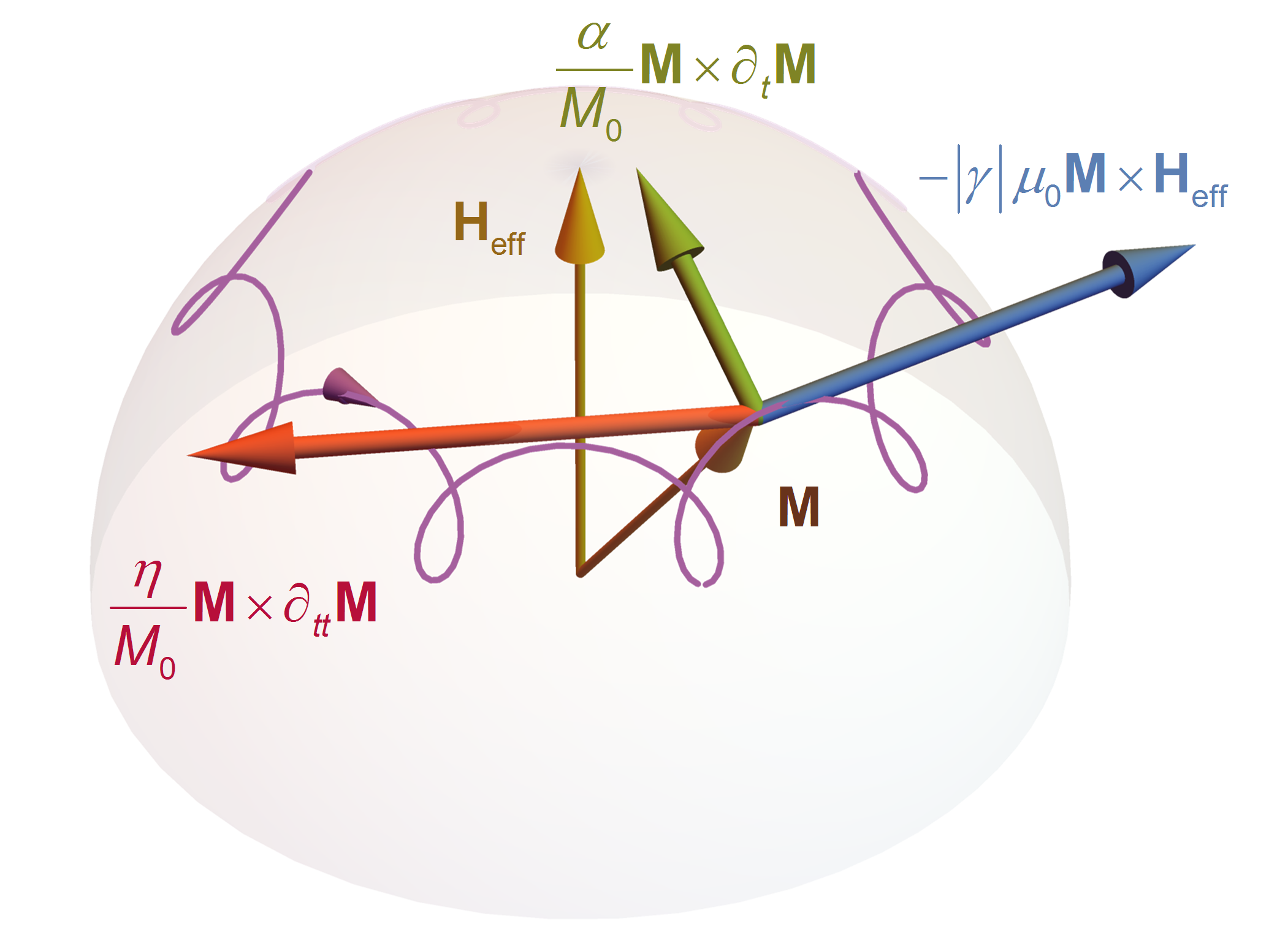}
    \caption{The magnetization ${\bf{M}}$ precesses around the effective magnetic field and ${{\bf{H}}_{{\rm{eff}}}}$ due to the precessional torque (blue). The inertial torque (red) causes the magnetization to undergo a concurrent nutational motion. The total magnetization motion is subject to the damping torque (green).
    %The orientation of vectors representing torques in the ILLG equation and schematic representation of magnetization trajectory. Magnetization ${\bf{M}}$ (shown by dark red arrow) is precessing around effective magnetic field ${{\bf{H}}_{{\rm{eff}}}}$ (yellow) driven by the field torque (blue). Direction of the damping torque (green) is given by a vector product of magnetization and its time derivative. The torque corresponding to the inertia of magnetization (red) is proportional to the second time derivative of magnetization and is generally non-collinear with the field torque. It leads to the modification of main precession’s frequency as well as to an additional nutational motion. 
    }
    \label{fig:NutSchematic}
\end{figure}

Inertia leads mainly to nutation – a terahertz-frequency motion of magnetization superimposed on the regular gigahertz-frequency precession \cite{Olive_Lansac_Meyer_Hayoun_Wegrowe_2015} (Fig. \ref{fig:NutSchematic}). Nutational resonances have been discussed in ferromagnets \cite{Cherkasskii_Farle_Semisalova_2020} and antiferromagnets \cite{Mondal_Grossenbach_Rozsa_Nowak_2021,Mondal_2021,Mondal_Oppeneer_2021,Mondal2021SpinCurrent}. Moreover, traveling nutational spin waves \cite{Makhfudz_Olive_Nicolis_2020,Cherkasskii_Farle_Semisalova_2021,Lomonosov_Temnov_Wegrowe_2021} have been proposed. Besides nutational motion, inertia has been found to result in a frequency shift of the uniform magnetization precession \cite{Olive_Lansac_Meyer_Hayoun_Wegrowe_2015, Li2015_PhysRevB.92.140413} and spin waves \cite{Cherkasskii_Farle_Semisalova_2021} at gigahertz frequencies. Previous studies of inertial spin dynamics treated various ferromagnetic systems including nanoparticles and  nanostructures  \cite{Titov_Coffey_Kalmykov_Zarifakis_Titov_2021,Titov_Coffey_Kalmykov_Zarifakis_2021,Rahman_Bandyopadhyay_2021}; however, a general approach based on ILLG for a ferromagnet with an arbitrary magnetic anisotropy has not yet been proposed.

Here, we develop a holistic theoretical framework for a ferromagnet with an arbitrary magnetic anisotropy energy landscape in analogy to the Smit-Beljers (SB) approach. For the last six decades \cite{VONSOVSKII19661}, the Smit-Beljers formalism has been an indispensable tool for routinely predicting and analyzing macrospin ferromagnetic resonance and, with extensions, spin-wave resonances. Now, our theoretical framework allows for deriving the frequencies of the nutational and precessional resonances of anisotropic ferromagnets in the presence of inertia. We formulate the secular equation of inertial spin dynamics, provide its exact and approximate solutions, and discuss the interplay of inertia and magnetic anisotropy.

\section{Precessional and nutational resonance frequencies}

The Smit-Beljers approach was initially developed via linearizing the Landau-Lifshitz equation in spherical coordinates and using small-angle approximation around the equilibrium direction of magnetization \cite{smit1955ferromagnetic, suhl1955ferromagnetic}. Employing the periodic solution ansatz, the ferromagnetic resonance (FMR) frequency was derived as
	\begin{equation}  \label{eq:SB}
	\omega^2_{\rm{SB}}= \frac{|\gamma|^2\left(1+\alpha^2\right)}{{M_0^2{{\sin }^2}{\theta_0}}}\left( {{\partial _{\theta \theta }}F{\partial _{\phi \phi }}F - {{\left( {{\partial _{\theta \phi }}F} \right)}^2}} \right),
	\end{equation} 	
where ${\theta _0}$ is the equilibrium polar angle of magnetization. The equation gives a closed-form relation between the FMR frequency and magnetic anisotropy energy $F$ and allows for efficient and convenient numerical analysis of experimental data \cite{Farle1998ferromagnetic}.

In our approach, we similarly write the ILLG equation in spherical coordinates which now contains first and second derivatives of the angles. As detailed in Appendix\,\ref{suppl1}, the relative magnitude of the terms in this resulting equation can be analyzed by their prefactors expressed in the orders of the Gilbert damping parameter $\alpha$. Since the latter is typically $10^{-3}-10^{-2}$, we omit higher-order terms and arrive at 
    \begin{equation} \label{eq:ILLGsph_approx1_main} 
    \left\{ \begin{array}{l}
    {\partial _{tt}}\theta  = \dfrac{{\left| \gamma  \right|{\mu _0}{H_\theta }}}{\eta } - \dfrac{{\alpha \left| \gamma  \right|{\mu _0}{H_\varphi }}}{\eta } 
     + \dfrac{{{\partial _t}\varphi \sin \theta }}{\eta } \\ 
    + {\left( {{\partial _t}\varphi } \right)^2}\sin \theta \cos \theta ,
    \\
    
    {\partial _{tt}}\varphi \sin \theta  = \dfrac{{\left| \gamma  \right|{\mu _0}{H_\varphi }}}{\eta } + \dfrac{{\alpha \left| \gamma  \right|{\mu _0}{H_\theta }}}{\eta } 
    - \dfrac{{{\partial _t}\theta }}{\eta } \\ 
    - 2{\partial _t}\varphi {\partial _t}\theta \cos \theta .
    \end{array} 
    \right.
    \end{equation}

Using the small-angle approximation, we develop the equations around the equilibrium direction of magnetization which introduces second-order derivatives of the energy $F$. The system of equations can be further linearized employing a Jacobian matrix of the angles (Appendix \ref{suppl1}). Using the periodic solution ansatz, we arrive at a fourth-order characteristic polynomial constituting the secular equation of the inertial spin system:
    \begin{equation} \label{eq:SBC}
    \begin{split}
		& \left[ \frac{\omega ^2}{\left| \gamma  \right|^2} - \frac{{\left( {1 + {\alpha ^2}} \right)}}{{M_0^2{{\sin }^2}{\theta _0}}}\left( {{\partial _{\theta \theta }}F{\partial _{\phi \phi }}F - {{\left( {{\partial _{\theta \phi }}F} \right)}^2}} \right) \right] \\
		& - {\eta ^2 \omega ^2} \left[ \frac{{\omega ^2}}{\left| \gamma  \right|^2} - \frac{1}{\eta \left| \gamma  \right| M_0}\left( {{\partial _{\theta \theta }}F + \frac{{{\partial _{\varphi \varphi }}F}}{{{{\sin }^2}{\theta _0}}}} \right) \right]\\
		& - i\omega \frac{{\alpha}}{\left| \gamma  \right| M_0}\left( {{\partial _{\theta \theta }}F + \frac{{{\partial _{\varphi \varphi }}F}}{{{{\sin }^2}{\theta _0}}}} \right) = 0. 
	\end{split}
    \end{equation}
The first group of terms corresponds to Eq.\,\ref{eq:SB}. The second group of terms introduces inertia of magnetization. The third group of terms corresponds to the frequency-domain linewidth of the ferromagnetic resonance 
    \begin{equation} \label{eq:dw_SBC}
	    \Delta\omega_{\rm{SB}} = \frac{\left|\gamma  \right| \alpha}{M_0}\left( {{\partial _{\theta \theta }}F + \frac{{{\partial _{\varphi \varphi }}F}}{{{{\sin }^2}{\theta _0}}}} \right)
    \end{equation}

as it does in the noninertial case \cite{VONSOVSKII19661,Skrotskii_Kurbatov_1959}. The presented approach has the advantage to converge to the Smit-Beljers secular equation when the inertial parameter vanishes and can be written as
    \begin{equation} \label{eq:SBC_short}
		\left(\omega^2 - \omega_{\mathrm{SB}}^2\right) - \eta^2\omega^2\left( \omega^2 - \frac{1}{\eta\alpha}\Delta\omega_{\rm{SB}} \right) - i \omega\Delta\omega_{\rm{SB}} = 0
    \end{equation}
%By employing Eq.~(\ref{eq:SBC}) we find that inertia shifts the FMR frequency down in the entire spectral range and the magnitude of the shift increases with increasing frequency. This effect can be well understood if one considers that the torque responsible for inertia is non-collinear and almost opposite to the field torque responsible for conventional precession (Fig. \ref{fig:NutSchematic})
Equation~\ref{eq:SBC_short} has two physical solutions: precessional resonance $\omega_{\mathrm{p}}$ at lower frequency and nutational resonance $\omega_{\mathrm{n}}$ at higher frequency. In Appendix\,\ref{suppl2}, we calculate the explicit (exact but complex) solutions, shown in Fig.\,2, as a benchmark for the consecutive approximations.   

First, similarly to the original Smit-Beljers formalism, we can omit the imaginary term in the secular Equation~\ref{eq:SBC_short} -- an approximation that we mark with 'a' -- and derive an analytical form of the resonance frequencies:
\begin{equation} \label{eq:w_approx_a_prec_main}
	\omega _{\rm{p}}^{\left( {\rm{a}} \right)} =  {\left( { p - \sqrt {{{p}^2} - q} } \right)^{1/2}},
\end{equation}
\begin{equation} \label{eq:w_approx_a_nut_main}
	\omega _{\rm{n}}^{\left( {\rm{a}} \right)} = {\left( { p + \sqrt {{{p}^2} - q} } \right)^{1/2}},
	\end{equation}
where
    \begin{equation} \label{eq:pq_main}
	\begin{array}{l}
    p =  \dfrac{1}{{{2 \eta ^2}}} + \dfrac{\Delta\omega_{\rm{SB}}}{2 \alpha\eta},\\
    q = \dfrac{\omega_{\rm{SB}}^2}{\eta^2}.
    \end{array}
    \end{equation}
The analytical form can be used conveniently to calculate the resonance frequencies via $\omega_{\rm{SB}}$ and $\Delta\omega_{\rm{SB}}$, thus adding just a few extra steps compared to the Smit-Beljers formalism. However, the analytical form is still too complex and the effect of magnetic anisotropy on precessional and nutational behavior is not immediately clear. 

\begin{figure*}[t]
    \centering
    \includegraphics[width=1\textwidth]{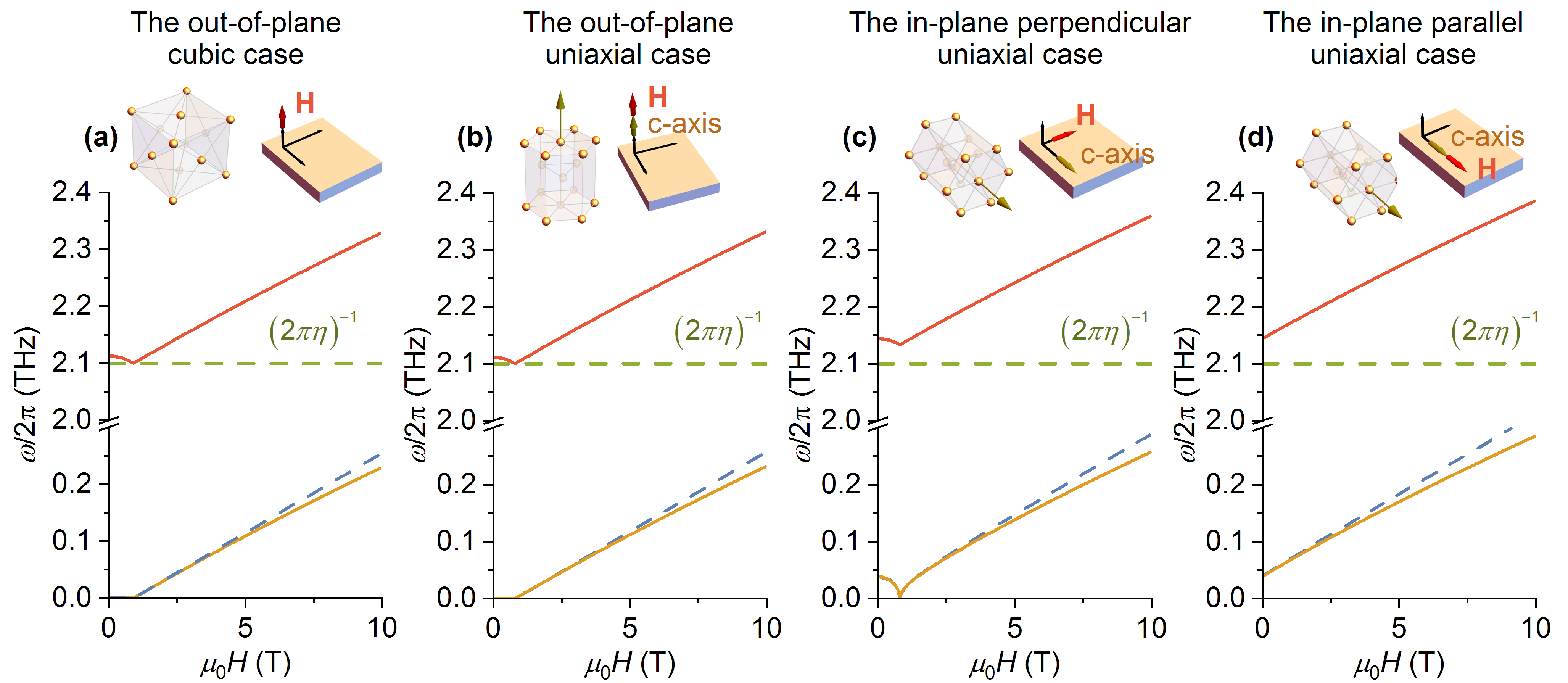}
    \caption{Frequency-field relation of ferromagnetic resonance and nutational resonances. Explicit solutions of Eq.\,\ref{eq:SBC_short} for the precessional resonance (orange) show a  red-shift compared to the non-inertial Smit-Beljers case (dashed blue). Explicit solutions of Eq.\,\ref{eq:SBC_short} for the nutational resonance (red) show a blue-shift compared to the zeroth-order approximation $1/\eta$. (a)~The calculation parameters for a thin film with cubic magnetocrystalline anisotropy are: $g = 2.09$, ${\mu _0}{M_0} =2.1 \:\rm{T}$, $\alpha=0.002$, $\eta = 75 \:\rm{fs} \cdot \rm{rad}^{-1}$, $K_{\rm{cub1}} =  4.9 \times 10^4 \: \rm{ J} \cdot \rm{m}^{-3}$, Ref. \cite{goryunov1995magnetic, Neeraj_Awari2021, Unikandanunni2021}.  (b)-(d)~The following parameters for a thin film with uniaxial magnetocrystalline anisotropy have been used: $g = 2.17$, ${\mu _0}{M_0} = 1.8 \: \rm{T}$, $\alpha  = 0.10$, $\eta = 75 \: \rm{fs} \cdot \rm{rad}^{ - 1}$, $K_{\rm{u1}} = 4.1 \times {10^5} \: \rm{J} \cdot \rm{m}^{ - 3}$, Ref. \cite{coey2010magnetism, Neeraj_Awari2021, Unikandanunni2021}.}
    \label{fig:FreqH}
\end{figure*}

We thus implement another approximation -- marked with 'b' -- by expanding the analytical form into a Taylor series (Appendix\,\ref{suppl2}) and neglecting higher-order terms in $\Delta\omega_{\rm{SB}}\eta\ll1$:
    \begin{equation} \label{eq:w_approx_b_prec_main}
    \begin{split}
    {\omega_{\rm{p}}^{(\rm{b})}} = \omega_{\rm{SB}} \sqrt{1-\eta\frac{\Delta\omega_{\rm{SB}}}{\alpha}},
    \end{split}
    \end{equation}
    \begin{equation} \label{eq:w_approx_b_nut_main}
	{\omega} _{\rm{n}}^{(\rm{b})} = \frac{1}{\eta } + \frac{\Delta\omega_{\rm{SB}}}{2\alpha}.
	\end{equation}
Here, we immediately see a systematic red-shift of the precessional frequency as compared to the non-inertial case of $\omega_{\rm{SB}}$. As shown in Fig.\,\ref{fig:NutSchematic}, the inertial torque vector has a notable component that is antiparallel to the precessional torque, thus effectively reducing the latter.

The nutational frequency ${\omega}_{\rm{n}}^{(\rm{b})}$, on the other hand, shows a substantial frequency increase (a blueshift) as compared to an earlier estimation $\omega_{\mathrm{n}}\sim1/\eta$ for an isotropic ferromagnet \cite{Neeraj_Awari2021,Unikandanunni2021}. Another approximation, for instance employed in Ref.\,\cite{Olive_Lansac_Meyer_Hayoun_Wegrowe_2015}, accounts for a frequency shift due to an applied magnetic field 
	\begin{equation} \label{eq:w_approx_Olive_nut}
	{\bar{\omega}_{\rm{n}}} = \frac{{\sqrt {1 + \eta \left| \gamma  \right|{\mu _0}{H_0}} }}{\eta } = \frac{1}{\eta } + \frac{{\left| \gamma  \right|{\mu _0}{H_0}}}{2} + ...
	\end{equation}
but neglects the effects of magnetic anisotropy. While the nutational frequency obtained in our model converges to the estimate $\bar{\omega}_{\rm{n}}$ in the case of vanishing anisotropy, it demonstrates that magnetic anisotropy (both, shape and magnetocrystalline) shifts the nutation resonance frequency as compared with  ${\bar{\omega}_{\rm{n}}}$, and must be accounted for according to the characteristic polynomial (Eq.\,\ref{eq:SBC_short}) and its solutions.

\section{Effect of magnetic anisotropy on inertial spin dynamics}

We calculate the effect of magnetic anisotropy on precessional and nutational resonances for four concrete examples of magnetic samples that have been and may likely be used in an experiment probing inertial spin dynamics. We consider single-crystal ferromagnetic thin films with cubic magnetocrystalline anisotropy (iron) and uniaxial magnetocrystalline anisotropy (hexagonal-close-packed cobalt) with magnetic parameters obtained from experimental data of Refs.\,\cite{Neeraj_Awari2021,Unikandanunni2021}. The explicit (exact) solutions of the characteristic polynomial of Eq.\,(\ref{eq:SBC_short}) are plotted in Fig.\,\ref{fig:FreqH} for various configurations of applied magnetic field with respect to the film surface and crystal symmetry axes. We use free energy density and equilibrium angles defined in Appendices~\ref{suppl3} and \ref{suppl4}. As shown in Fig.\,\ref{fig:FreqH}, the effect of inertia is consistent in all calculated scenarios. The precessional frequency experiences a redshift due to inertia as compared to resonance frequency $\omega_{\rm{SB}}$ for noninertial Smit-Beljers case. Both aligned precessional modes (above the saturation field) and nonaligned precessional modes (below the saturation field in hard-axes configurations) \cite{Farle1998ferromagnetic} experience a redshift which increases with increasing precessional frequency. For the aligned modes, the redshift thus becomes stronger with increasing magnetic field. For nonaligned modes, on the other hand, decreasing magnetic field can result in increasing redshift.

The nutational frequency experiences a blue shift due to magnetic anisotropy and magnetic field. The blueshift typically increases with increasing magnetic field. However, below the saturation field in hard-axes configurations, the blue-shift of the nutational frequency can become stronger with decreasing magnetic field (see the red line in Fig.\,\ref{fig:FreqH}(a-c)). The exact behavior is dominated by the $\Delta \omega_{\mathrm{SB}}$ term in Eqs.\,(\ref{eq:w_approx_b_prec_main}) and (\ref{eq:w_approx_b_nut_main}). Both shifts can be substantial (reaching up to 12\% in the magnetic field of 10T) for all calculated scenarios.

While we use the explicit solutions of the secular equation in Fig.\,\ref{fig:FreqH} to visualize the effects of magnetic anisotropy in inertial spin systems, the observed frequency shifts are in qualitative agreement with the approximations. To assess the quantitative validity of our approximations 'a' and 'b', we compare them with the benchmark of the explicit solutions. We find that the analytical form 'a' (Eqs.\,(\ref{eq:w_approx_a_prec_main}-\ref{eq:w_approx_a_nut_main})) accrues less than 0.5\% error for aligned modes when compared to the exact solution of Eq.\,\ref{eq:SBC}; however, it should be stressed that the characteristic polynomial Eq.\,\ref{eq:SBC} itself has been derived for inertial parameters $\eta \ll 1/\Delta\omega_{\rm{SB}}$. The next-step approximation by the Taylor series 'b' (Eqs.\,(\ref{eq:w_approx_b_prec_main}-\ref{eq:w_approx_b_nut_main})) introduces an error less than 6\% for aligned modes with $\eta < 100 \: \rm{fs} \cdot \rm{rad}^{ - 1}$, while at higher values of the inertial parameter, the Taylor series causes a substantial error of the frequencies. While the approximation 'b' in Eqs.\,(\ref{eq:w_approx_b_prec_main}-\ref{eq:w_approx_b_nut_main}) should thus be treated with caution, for comparison, we calculate the explicit forms of precessional and nutational frequencies for aligned modes in the configurations displayed in Fig.\,\ref{fig:FreqH}:

(a)~The out-of-plane cubic case:
	\begin{equation} \label{eq:w_approx_prec_b_cub_H_OOP}
    \begin{split}
    &{\omega_{\rm{p}}^{(\rm{b})}}^2 = 
    \left| \gamma  \right|^2\left(1+ \alpha ^2 \right) \left(-\mu_0 M_0 + \mu_0 H_0  + \dfrac{2 K_{\rm{cub1}}}{M} \right)^2 \\
    & \times \left[1-\eta\left| \gamma  \right|\left(-2 \mu_0 M_0 +2 \mu_0 H_0  + \dfrac{4 K_{\rm{cub1}}}{M}\right)\right],
    \end{split}
    \end{equation}

    \begin{equation} \label{eq:w_approx_b_nut_cub_H_OOP}
	{\omega}_{\rm{n}}^{(\rm{b})} = \frac{1}{\eta } + \left| \gamma  \right|\left( { - {\mu _0}{M_0} + {\mu _0}{H_0} + \frac{{2{K_{{\rm{cub1}}}}}}{{{M_0}}}} \right)
	\end{equation}

(b)~The out-of-plane uniaxial case:

    \begin{equation} \label{eq:w_approx_b_prec_uni_OOP_H_OOP}
    \begin{split}
    {\omega_{\rm{p}}^{(\rm{b})}}^2 = \left(1+\alpha ^2\right) | \gamma | ^2 \left( -\mu_0 M_0 + \mu_0 H_0 +\dfrac{2 K_{\rm{u1}}}{M_0}\right)^2 \times \\ 
    \left(1- \eta  | \gamma |  \left(-2\mu_0 M_0 + 2\mu_0 H_0 +\dfrac{4 K_{\text{u1}}}{M_0} \right)\right),
    \end{split}
    \end{equation}

    \begin{equation} \label{eq:w_approx_b_nut_uni_OOP_H_OOP}
	\omega _{\rm{n}}^{\left( {\rm{b}} \right)} = \frac{1}{\eta } + \left| \gamma  \right|\left( { - {\mu _0}{M_0} + {\mu _0}{H_0} + \frac{{2{K_{{\rm{u1}}}}}}{{{M_0}}}} \right).
	\end{equation}

(c)~The in-plane perpendicular uniaxial case:

    \begin{equation} \label{eq:w_approx_b_prec_uni_IP_H_IP_perp}
    \begin{split}
    \begin{array}{l}
    {\omega_{\rm{p}}^{(\rm{b})}}^2 = \left(1+ \alpha ^2\right) | \gamma | ^2 \left(\mu_0 M_0 + \mu_0 H_0 \right) \left(\mu_0 H_0 -\dfrac{2 K_{\rm{u1}}}{M_0}\right) \times \\
    \left(1-\eta  | \gamma |  \left(\mu_0 M_0 + 2 \mu_0 H_0 -\dfrac{2 K_{\rm{u1}}}{M_0}\right)\right),
    \end{array}
    \end{split}
    \end{equation}
    
    \begin{equation} \label{eq:w_approx_b_nut_uni_IP_H_IP_perp}
	\omega _{\rm{n}}^{\left( {\rm{b}} \right)} = \frac{1}{\eta } + \left| \gamma  \right|\left( {\frac{{{\mu _0}{M_0}}}{2} + {\mu _0}{H_0} - \frac{{{K_{{\rm{u1}}}}}}{{{M_0}}}} \right).
	\end{equation}

(d)~The in-plane parallel uniaxial case:

    \begin{equation} \label{eq:w_approx_b_prec_uni_IP_H_IP_par}
    \begin{split}
    \begin{array}{l}
    {\omega_{\rm{p}}^{(\rm{b})}}^2=\left(1+ \alpha^2 \right) | \gamma |^2 \left(\mu_0 H_0 +\dfrac{2 K_{\rm{u1}}}{M_0}\right) \times \\ 
    \left(\mu_0 M_0 + \mu_0 H_0 +\dfrac{2 K_{\rm{u1}}}{M_0}\right) \times \\ 
    \left(1-\eta  | \gamma |  \left(\mu_0 M_0 + 2 \mu_0 H_0 +\dfrac{4 K_{\rm{u1}}}{M_0}\right)\right),
    \end{array}
    \end{split}
    \end{equation}
    
    \begin{equation}  \label{eq:w_approx_b_nut_uni_IP_H_IP_par}
	\omega _{\rm{n}}^{\left( {\rm{b}} \right)} = \frac{1}{\eta } + \left| \gamma  \right|\left( {\frac{{{\mu _0}{M_0}}}{2} + {\mu _0}{H_0} + \frac{{2{K_{{\rm{u1}}}}}}{{{M_0}}}} \right).
	\end{equation} 	
	
It is a common practice to analyze experimentally determined dependences of ferromagnetic resonance frequency on applied magnetic field for evaluating magnetic parameters such as magnetic anisotropy and g-factor \cite{Finizio.PhysRevB.98.104415,OrtizMAEEIG,Farle1998ferromagnetic,gonccalves2018oscillatory,Etesamirad3MSwitch,Barsukov3M,Navabi3M}. Our work demonstrates that such evaluation needs to be adjusted by taking into account the inertial redshift. In particular, measurements at higher fields/frequencies have been considered to result in more accurate determination of magnetic parameters \cite{ShawGFactor,BarsukovFieldDepMAE,BarsukovMagneticPhase,Ordonez-Romero_Cherkasskii_Qureshi_Kalinikos_Patton_2013,Wang_Cherkasskii_Kalinikos_Carr_Wu_2014}. Our model, however, shows that especially at high magnetic fields, inertial redshift is strong and needs to be taken into account. %{\blue At low magnetic fields (XXX or at low frequencies ???), on the other hand, the redshift is found to be small.}

It should be noted that in the framework of the extended breathing Fermi surface model \cite{Fahnle_Steiauf_Illg_2011, fahnle2013erratum}, the inertial term with negative sign was derived. Such negative inertial term would formally result in a blueshift of the precessional frequencies. However, since the origin of inertia is still under discussion, we consider here only the effects of the positive inertial term suggested in Ref.\,\cite{Wegrowe_Ciornei_2012}.

%Overall, we find that with the inertia-induced redshift of the ferromagnetic resonance and a higher than $\eta^{-1}$ boundary nutational resonance, crossing of these two frequencies in the frequency-field diagram (Fig. \ref{fig:FreqH}) is not possible.

\section{Summary}

In summary, we derived the secular equation for an inertial spin system with an arbitrary magnetic anisotropy energy in analogy with the Smit-Beljers approach. We find that ferromagnetic resonance experiences a substantial redshift due to the inertia, while nutational resonance experiences a blueshift due to magnetic anisotropy and field. For an accurate evaluation of magnetic parameters from magnetic resonance measurements, inertia needs to be taken into account. Our model (Eq.\,\ref{eq:SBC_short}) allows for convenient calculation of precessional and nutational resonances of an inertial spin system using parameters ($\omega_{\mathrm{SB}}$ and $\Delta\omega_{\mathrm{SB}}$) obtained from noninertial models.

%Since inertia affects magnetization dynamics throughout the entire gigahertz and terahertz frequency range, previous investigation results in linear and nonlinear spin dynamics \cite{Cherkasskii_Nikitin_Kalinikos_2016, Ordonez-Romero_Cherkasskii_Qureshi_Kalinikos_Patton_2013, Wang_Cherkasskii_Kalinikos_Carr_Wu_2014} should probably be interpreted differently. Note that here we presented linear theory, which can be extended by including the description of high-order nutation resonances, which were detected recently \cite{Unikandanunni2021}.

% \section{DATA AVAILABILITY}
% The data that support the findings of this study are available from the corresponding authors upon reasonable request.
% \section{CODE AVAILABILITY}
% The code used for simulations is available from the corresponding authors upon reasonable request.
% Received: xxxxxx; Accepted: xxxxxx; Published online: xxxxxx.
% \section{REFERENCES}
% Will be moved here
\section*{ACKNOWLEDGMENTS}
M.Ch., M.F. and A.S. acknowledge funding by Deutsche Forschungsgemeinschaft (DFG, German Research Foundation) – Project No. 392402498 (SE 2853/1-1) and Project No. 405553726 CRC/TRR 270. R.M. acknowledges the funding from the Swedish Research Council via VR 2019-06313. I.B. acknowledges funding by the National Science Foundation through Grant No. ECCS-1810541.
% \section{AUTHOR CONTRIBUTIONS}
% M.Ch., M.F., A.S., R.M., and I.B. contributed equally.
% \section{COMPETING INTERESTS}
% The authors declare no competing interests.

\appendix
\section{SMIT-BELJERS APPROACH WITH INERTIA}
\label{suppl1}

First, we transform the ILLG equation into a spherical coordinate system and find equilibrium angles of magnetization. Second, we linearize the system of equations describing magnetization dynamics at the equilibrium point. Finally, we derive the eigenfrequencies corresponding to the resonances.

In a spherical coordinate system one writes ${\bf{M}} = {M_0}{{\bf{e}}_{\rm{r}}},$ ${{\bf{H}}_{{\rm{eff}}}} = {H_{\rm{r}}}{{\bf{e}}_{\rm{r}}} + {H_{\rm{\theta }}}{{\bf{e}}_{\rm{\theta }}} + {H_{\rm{\varphi }}}{{\bf{e}}_{\rm{\varphi }}},$ where the magnitude of the magnetization vector persists over time. Using
    \begin{equation} \label{eq:diff_tt_M}
	\begin{array}{l}
    {\partial _t}{\bf{M}} = {M_0}\left( {{\partial _t}\theta {{\bf{e}}_{\rm{\theta }}} + \sin \theta {\partial _t}\varphi {{\bf{e}}_{\rm{\varphi }}}} \right),\\
    {\partial _{tt}}{\bf{M}} = {M_0}\left\{ {\left[ { - {{\left( {{\partial _t}\theta } \right)}^2} - {{\left( {{\partial _t}\varphi } \right)}^2}{{\sin }^2}\theta } \right]{{\bf{e}}_{\rm{r}}}} \right. \\[4pt]
    + \left[ {{\partial _{tt}}\theta  - {{\left( {{\partial _t}\varphi } \right)}^2}\sin \theta \cos \theta } \right]{{\bf{e}}_{\rm{\theta }}} \\[5pt] \left. { + \left[ {{\partial _{tt}}\varphi \sin \theta  + 2{\partial _t}\varphi {\partial _t}\theta \cos \theta } \right]{{\bf{e}}_{\rm{\varphi }}}} \right\},
    \end{array}
	\end{equation}
one transforms the ILLG equation into
    \begin{equation} \label{eq:ILLGsph}
	\left\{ \begin{array}{l}
    {\partial _{tt}}\theta  = \dfrac{{\left| \gamma  \right|{\mu _0}{H_{\rm{\theta }}}}}{\eta } - \dfrac{{\alpha \textcolor{goodred}{{\partial _t}\theta} }}{\eta } + \dfrac{{{\partial _t}\varphi \sin \theta }}{\eta } \\ + {\left( {{\partial _t}\varphi } \right)^2}\sin \theta \cos \theta ,
    \\[8pt]
    {\partial _{tt}}\varphi \sin \theta  = \dfrac{{\left| \gamma  \right|{\mu _0}{H_{\rm{\varphi }}}}}{\eta } - \dfrac{{\alpha \textcolor{goodred}{{\partial _t}\varphi \sin \theta} }}{\eta } - \dfrac{{{\partial _t}\theta }}{\eta } \\ - 2{\partial _t}\varphi {\partial _t}\theta \cos \theta .
    \end{array} \right.
    \end{equation}
Here, we introduce the first approximation, i.e. we replace the "red" terms in the system (\ref{eq:ILLGsph}) with the "blue" ones in the system (\ref{eq:ILLGsph_approx1}) as follows. Based on the ILLG equation, we write
    \begin{equation} \label{eq:ILLGsph_reshape1}
    \left\{ \begin{array}{l}
    \dfrac{{\alpha {\partial _t}\varphi \sin \theta }}{\eta } =  - \dfrac{{\alpha \left| \gamma  \right|{\mu _0}{H_\theta }}}{\eta } + \dfrac{{{\alpha ^2}{\partial _t}\theta }}{\eta }\\
    {\rm{   }} - \alpha {\left( {{\partial _t}\varphi } \right)^2}\sin \theta \cos \theta  + \alpha {\partial _{tt}}\theta,\\[8pt]
    
    \dfrac{{\alpha {\partial _t}\theta }}{\eta } = \dfrac{{\alpha \left| \gamma  \right|{\mu _0}{H_\varphi }}}{\eta } - \dfrac{{{\alpha ^2}{\partial _t}\varphi \sin \theta }}{\eta }\\
    {\rm{   }} - 2\alpha {\partial _t}\varphi {\partial _t}\theta \cos \theta  - \alpha {\partial _{tt}}\varphi \sin \theta
    
    \end{array} \right.
    \end{equation}
and substitute these equations in (\ref{eq:ILLGsph}) instead of the corresponding "color" terms. We obtain
    \begin{equation} \label{eq:ILLGsph_with_reshape1}
    \left\{ \begin{array}{l}
    {\partial _{tt}}\theta  = \dfrac{{\left| \gamma  \right|{\mu _0}{H_\theta }}}{\eta } - \dfrac{{\alpha \left| \gamma  \right|{\mu _0}{H_\varphi }}}{\eta } + \dfrac{{{\partial _t}\varphi \sin \theta }}{\eta }\left( {1 + {\alpha ^2}} \right) \\ + {\left( {{\partial _t}\varphi } \right)^2}\sin \theta \cos \theta
    {\rm{  }} + 2\alpha {\partial _t}\varphi {\partial _t}\theta \cos \theta  + \alpha {\partial _{tt}}\varphi \sin \theta ,\\[8pt]
    {\partial _{tt}}\varphi \sin \theta  = \dfrac{{\left| \gamma  \right|{\mu _0}{H_\varphi }}}{\eta } + \dfrac{{\alpha \left| \gamma  \right|{\mu _0}{H_\theta }}}{\eta } - \dfrac{{{\partial _t}\theta }}{\eta }\left( {1 + {\alpha ^2}} \right) \\  - 2{\partial _t}\varphi {\partial _t}\theta \cos \theta
    {\rm{  }} + \alpha {\left( {{\partial _t}\varphi } \right)^2}\sin \theta \cos \theta  - \alpha {\partial _{tt}}\theta .
    \end{array} \right.
    \end{equation}
The last two terms in both equations are negligible, since they are multiplied by $\alpha < 1$, while $\Delta\omega_{\rm{SB}}\eta  \ll 1$. The ${\alpha ^2}$ terms are much less than 1, hence the ILLG equation is converted to
    \begin{equation} \label{eq:ILLGsph_approx1}
	\left\{ \begin{array}{l}
    {\partial _{tt}}\theta  = \dfrac{{\left| \gamma  \right|{\mu _0}{H_{\rm{\theta }}}}}{\eta } - \dfrac{{\alpha \textcolor{goodblue}{\left| \gamma  \right|{\mu _0}{H_{\rm{\varphi }}}}}}{\eta } + \dfrac{{{\partial _t}\varphi \sin \theta }}{\eta } \\ + {\left( {{\partial _t}\varphi } \right)^2}\sin \theta \cos \theta ,
    \\[8pt]
    {\partial _{tt}}\varphi \sin \theta  = \dfrac{{\left| \gamma  \right|{\mu _0}{H_{\rm{\varphi }}}}}{\eta } + \dfrac{{\alpha \textcolor{goodblue}{\left| \gamma  \right|{\mu _0}{H_{\rm{\theta }}}}}}{\eta } - \dfrac{{{\partial _t}\theta }}{\eta } \\ - 2{\partial _t}\varphi {\partial _t}\theta \cos \theta .
    \end{array} \right.
    \end{equation}
This transformation is commonly adopted and was performed in Ref.\,\cite{Skrotskii_Kurbatov_1959}. The advantage of the first approximation is that the final result, which is to be shown below, converges to the SB equation for $\eta  = 0$.
Note that the effective magnetic field ${{\bf{H}}_{{\rm{eff}}}} =  - \mu _0^{ - 1}{\partial _{\bf{M}}}F$ in the spherical coordinate system is given by
    \begin{equation} \label{eq:Heff_derF}
	{H_{\rm{\theta }}} =  - \frac{1}{{{\mu _0}{M_0}}}{\partial _\theta }F,
	\,
	{H_{\rm{\varphi }}} =  - \frac{1}{{{\mu _0}{M_0}\sin \theta }}{\partial _\varphi }F.
	\end{equation}

In order to find the eigenfrequencies from the nonlinear system of equations (\ref{eq:ILLGsph_approx1}), it is necessary to linearize it and to determine the equilibrium orientation of magnetization. The equilibrium given by the angles ${\theta _0}$ and ${\varphi _0}$ is found from the extremum conditions
    \begin{equation}	\label{eq:D_F=0} 
    {\partial _\theta }F = 0, \, {\partial _\varphi }F = 0
    \end{equation}
limited by the conditions for the minimum, namely, the determinant of a Hessian matrix has to be positive
	\begin{equation}	\label{eq:Hessian} 
	{\partial _{\theta \theta }}F{\partial _{\varphi \varphi }}F - {\partial _{\theta \varphi }}F{\partial _{\varphi \theta }}F > 0
	\end{equation}
and one of the second derivative has to be positive as well
    \begin{equation}	\label{eq:F_second_derivative_positive} 
	{\partial _{\theta \theta }}F > 0.
	\end{equation}
In the excited state, magnetization is deflected from the equilibrium orientation by the effective magnetic field changes over time. Here, we introduce the second approximation, which corresponds to the standard SB approach, that is the deflection from equilibrium is small
    \begin{equation}	\label{eq:deltaTheta_deltaPhi} 
	\Delta \theta \left( t \right) = \theta \left( t \right) - {\theta _0}
	,\,
	\Delta \varphi \left( t \right) = \varphi \left( t \right) - {\varphi _0}
	\end{equation}
and it is sufficient to limit the expansion of free energy to the linear terms 
    \begin{equation}	\label{eq:DF_expansion} 
	\begin{array}{l}
    {\partial _\theta }F = {\partial _{\theta \theta }}F\Delta \theta  + {\partial _{\theta \varphi }}F\Delta \varphi ,\\[4pt]
    {\partial _\varphi }F = {\partial _{\theta \varphi }}F\Delta \theta  + {\partial _{\varphi \varphi }}F\Delta \varphi,
    \end{array}
    \end{equation}
where the second derivatives are evaluated at the equilibrium. Using the small-angle approximation, one obtains from expressions (\ref{eq:ILLGsph_approx1})-(\ref{eq:DF_expansion})
    \begin{equation}	\label{eq:dTh_dPhi_with_DF_expansion}
	\begin{array}{l}
    {\partial _{tt}}\Delta \theta  = \left( { - \dfrac{{\left| \gamma  \right|}}{{\eta {M_0}}}{\partial _{\theta \theta }}F + \dfrac{{\alpha \left| \gamma  \right|}}{{\eta {M_0}\sin {\theta _0}}}{\partial _{\theta \varphi }}F} \right)\Delta \theta 
    \\[8pt] + \left( { - \dfrac{{\left| \gamma  \right|}}{{\eta {M_0}}}{\partial _{\theta \varphi }}F + \dfrac{{\alpha \left| \gamma  \right|}}{{\eta {M_0}\sin {\theta _0}}}{\partial _{\varphi \varphi }}F} \right)\Delta \varphi \\[8pt]
    + \dfrac{{{\partial _t}\Delta \varphi \sin {\theta _0}}}{\eta } + {\left( {{\partial _t}\Delta \varphi } \right)^2}\sin {\theta _0}\cos {\theta _0}
    ,\\[8pt]
    {\partial _{tt}}\Delta \varphi \sin {\theta _0} = \left( { - \dfrac{{\left| \gamma  \right|}}{{\eta {M_0}\sin {\theta _0}}}{\partial _{\theta \varphi }}F - \dfrac{{\alpha \left| \gamma  \right|}}{{\eta {M_0}}}{\partial _{\theta \theta }}F} \right)\Delta \theta \\[8pt]  + \left( { - \dfrac{{\left| \gamma  \right|}}{{\eta {M_0}\sin {\theta _0}}}{\partial _{\varphi \varphi }}F - \dfrac{{\alpha \left| \gamma  \right|}}{{\eta {M_0}}}{\partial _{\theta \varphi }}F} \right)\Delta \varphi \\[8pt]
    - \dfrac{{{\partial _t}\Delta \theta }}{\eta } - 2{\partial _t}\Delta \varphi {\partial _t}\Delta \theta \cos {\theta _0}.
    \end{array}
    \end{equation}

In order to linearize this system of equations, the following notations are introduced
    \begin{equation}	\label{eq:a_coeff}
	\begin{array}{l}
    {a_{41}} =  - \dfrac{{\left| \gamma  \right|{\partial _{\theta \varphi }}F}}{{\eta {M_0}{{\sin }^2}{\theta _0}}} - \dfrac{{\alpha \left| \gamma  \right|{\partial _{\theta \theta }}F}}{{\eta {M_0}\sin {\theta _0}}},\\[8pt]
    {a_{42}} =  - \dfrac{1}{{\eta \sin {\theta _0}}},\\[8pt]
    {a_{43}} =  - \dfrac{{\left| \gamma  \right|{\partial _{\varphi \varphi }}F}}{{\eta {M_0}{{\sin }^2}{\theta _0}}} - \dfrac{{\alpha \left| \gamma  \right|{\partial _{\theta \varphi }}F}}{{\eta {M_0}\sin {\theta _0}}},\\[8pt]
    {\nu _4} =  - 2\cot {\theta _0},\\[8pt]
    {a_{21}} = \dfrac{\alpha {\left| \gamma  \right|{\partial _{\theta \varphi }}F}}{{\eta {{M}_0}\sin {\theta _0}}} - \dfrac{{\left| \gamma  \right|{\partial _{\theta \theta }}F}}{{\eta {{M}_0}}},\\[8pt]
    {a_{23}} = \dfrac{{\alpha \left| \gamma  \right|{\partial _{\varphi \varphi }}F}}{{\eta {M_0}\sin {\theta _0}}} - \dfrac{{\left| \gamma  \right|{\partial _{\theta \varphi }}F}}{{\eta {M_0}}},\\[8pt]
    {a_{24}} = \dfrac{{\sin {\theta _0}}}{\eta },\\[8pt]
    {\nu _2} = \sin {\theta _0}\cos {\theta _0},\\[8pt]
    {x_1} = \Delta \theta ,\\[8pt]
    {x_2} = {\partial _t}\Delta \theta ,\\[8pt]
    {x_3} = \Delta \varphi ,\\[8pt]
    {x_4} = {\partial _t}\Delta \varphi .
    \end{array}
    \end{equation}
Employing (\ref{eq:a_coeff}), we rewrite the system (\ref{eq:dTh_dPhi_with_DF_expansion}) as
	\begin{equation}	\label{eq:eq_x}
    \begin{array}{l}
    {\partial _t}{x_1} = {x_2},\\
    {\partial _t}{x_2} = {a_{21}}{x_1} + {a_{23}}{x_3} + {a_{24}}{x_4} + {\nu _2}x_4^2,\\
    {\partial _t}{x_3} = {x_4},\\
    {\partial _t}{x_4} = {a_{41}}{x_1} + {a_{42}}{x_2} + {a_{43}}{x_3} + {\nu _4}{x_2}{x_4}.
    \end{array}
    \end{equation}
At the fixed point ${{\bf{x}}^*} = \left( {x_1^*,x_2^*,x_3^*,x_4^*} \right)$ the dynamics of the nonlinear system (\ref{eq:eq_x}) are qualitatively similar to the dynamics of a linear system (\ref{eq:general_lin_eq}) associated with the Jacobian matrix $J\left( {{{\bf{x}}^*}} \right)$ \cite{Verhulst_1996}, i.e.,
    \begin{equation}	\label{eq:general_lin_eq}
    \begin{split}
	& \left( {\begin{array}{*{20}{l}}
    {\partial _t}{x_1}\\
    {\partial _t}{x_2}\\
    {\partial _t}{x_3}\\
    {{\partial _t}{x_4}}
    \end{array}} \right) = \\
    & \left( {\begin{array}{*{20}{c}}
    {{\partial _{{x_1}}}{f_1}\left( {{{\bf{x}}^*}} \right)}& \ldots &{{\partial _{{x_4}}}{f_1}\left( {{{\bf{x}}^*}} \right)}\\
    \vdots & \ddots & \vdots \\
    {{\partial _{{x_1}}}{f_4}\left( {{{\bf{x}}^*}} \right)}& \ldots &{{\partial _{{x_4}}}{f_4}\left( {{{\bf{x}}^*}} \right)}
    \end{array}} \right)\left( {\begin{array}{*{20}{l}}
    {x_1}\\
    {x_2}\\
    {x_3}\\
    {x_4}
    \end{array}} \right),
    \end{split}
    \end{equation}
where the right-hand sides of Eqs. (\ref{eq:eq_x}) are denoted as ${f_i}$. The fixed point is determined by equating the derivatives of the nonlinear system (\ref{eq:eq_x}) to zero, which gives the following equations
    \begin{equation}	\label{eq:eq_x_eq2} 
	\begin{array}{l}
    {x_2} = 0,\\
    {a_{21}}{x_1} + {a_{23}}{x_3} = 0,\\
    {x_4} = 0,\\
    {a_{41}}{x_1} + {a_{43}}{x_3} = 0,
    \end{array}
    \end{equation}
with the solution $x_1^* = x_2^* = x_3^* = x_4^* = 0$. The Jacobian matrix of Eqs. (\ref{eq:eq_x}) is
    \begin{equation}	\label{eq:Jacobian_matrix} 
	J = \left[ {\begin{array}{*{20}{c}}
    0&1&0&0\\
    {{a_{21}}}&0&{{a_{23}}}&{{a_{24}} + 2{\nu _2}{x_4}}\\
    0&0&0&1\\
    {{a_{41}}}&{{a_{42}} + {\nu _4}{x_4}}&{{a_{43}}}&{{\nu _4}{x_2}}
    \end{array}} \right]
    \end{equation}
and at the point $x_1^* = x_2^* = x_3^* = x_4^* = 0$ it provides the linear system of equations
	\begin{equation}	\label{eq:x_eq_lin} 
    \begin{array}{l}
    {\partial _t}{x_1} = {x_2},\\
    {\partial _t}{x_2} = {a_{21}}{x_1} + {a_{23}}{x_3} + {a_{24}}{x_4},\\
    {\partial _t}{x_3} = {x_4},\\
    {\partial _t}{x_4} = {a_{41}}{x_1} + {a_{42}}{x_2} + {a_{43}}{x_3}.
    \end{array}
    \end{equation}
This third approximation goes beyond the SB approach and it is the linearization of the system (\ref{eq:eq_x}). The eigenvalues of these equations give resonance frequencies, which are calculated from the characteristic polynomial
    \begin{equation}	\label{eq:characteristic_equation}
    \begin{split}
	&{\omega ^4} + \left( {{a_{21}} + {a_{24}}{a_{42}} + {a_{43}}} \right){\omega ^2} \\ 
	& - i\left( {{a_{24}}{a_{41}} + {a_{23}}{a_{42}}} \right)\omega - {a_{23}}{a_{41}} + {a_{21}}{a_{43}} = 0.
	\end{split}
	\end{equation}
Restoring the original variable notations, one finds the equation describing eigenfrequencies of a ferromagnet with inertia
    \begin{equation} \label{eq:SBC_appendix}
    \begin{split}
		& \left[ \frac{\omega ^2}{\left| \gamma  \right|^2} - \frac{{\left( {1 + {\alpha ^2}} \right)}}{{M_0^2{{\sin }^2}{\theta _0}}}\left( {{\partial _{\theta \theta }}F{\partial _{\phi \phi }}F - {{\left( {{\partial _{\theta \phi }}F} \right)}^2}} \right) \right] \\
		& - {\eta ^2 \omega ^2} \left[ \frac{{\omega ^2}}{\left| \gamma  \right|^2} - \frac{1}{\eta \left| \gamma  \right| M_0}\left( {{\partial _{\theta \theta }}F + \frac{{{\partial _{\varphi \varphi }}F}}{{{{\sin }^2}{\theta _0}}}} \right) \right]\\
		& - i\omega \frac{{\alpha}}{\left| \gamma  \right| M_0}\left( {{\partial _{\theta \theta }}F + \frac{{{\partial _{\varphi \varphi }}F}}{{{{\sin }^2}{\theta _0}}}} \right) = 0. 
	\end{split}
	 \end{equation}
Note that this equation can be converted to SB formula (\ref{eq:SB}) if the inertial parameter vanishes.

\section{EXACT AND APPROXIMATE EXPRESSIONS OF RESONANCE FREQUENCIES}
\label{suppl2}

The quartic equation (\ref{eq:SBC_appendix}) results in two pairs of roots. The first pair is precessional frequency modified by inertia, one root of the pair is positive, the second is negative. The same applies to the other pair corresponding to the nutational frequency. Here we consider only positive roots. Let us use the Ferrari's solution for this quartic equation to write exact expressions of resonance frequencies, and introduce the notations:
    \begin{equation} \label{eq:Ac_coeff}
	\begin{split}
    &{A_{\rm{r}}} = M_0^2{\eta ^2},\\[4pt]
    &{C_{\rm{r}}} =  - M_0^2 - {M_0}\eta \left| \gamma  \right|\left( {{\partial _{\theta \theta }}F + \dfrac{{{\partial _{\varphi \varphi }}F}}{{{{\sin }^2}{\theta _0}}}} \right),\\[4pt]
    &{D_{\rm{r}}} = i{M_0}\alpha \left| \gamma  \right|\left( {{\partial _{\theta \theta }}F + \dfrac{{{\partial _{\varphi \varphi }}F}}{{{{\sin }^2}{\theta _0}}}} \right)\\[4pt]
    &{E_{\rm{r}}} = \dfrac{{{{\left| \gamma  \right|}^2}\left( {1 + {\alpha ^2}} \right)}}{{{{\sin }^2}{\theta _0}}}\left( {{\partial _{\theta \theta }}F{\partial _{\phi \phi }}F - {{\left( {{\partial _{\theta \phi }}F} \right)}^2}} \right),\\[4pt]
    &{a_{\rm{r}}} = \dfrac{{{C_{\rm{r}}}}}{{{A_{\rm{r}}}}},\; {b_{\rm{r}}} = \dfrac{{{D_{\rm{r}}}}}{{{A_{\rm{r}}}}},\; {c_{\rm{r}}} = \dfrac{{{E_{\rm{r}}}}}{{{A_{\rm{r}}}}}.
    \end{split}
    \end{equation}
In Ferrari's method, one determines a root of the nested depressed cubic equation. In our case, the root is written
    \begin{equation} \label{eq:yr_coeff}
	{y_{\rm{r}}} =  - \frac{{5{a_{\rm{r}}}}}{6} + {U_{\rm{r}}} + {V_{\rm{r}}},
	\end{equation}
where
    \begin{equation} \label{eq:UQ_coeff}
	\begin{array}{l}
    {U_{\rm{r}}} = \sqrt[3]{{ - \sqrt {\dfrac{{P_{\rm{r}}^3}}{{27}} + \dfrac{{Q_{\rm{r}}^2}}{4}}  - \dfrac{{{Q_{\rm{r}}}}}{2}}},\\
    {V_{\rm{r}}} =  - \dfrac{{{P_{\rm{r}}}}}{{3{U_{\rm{r}}}}},\\
    {P_{\rm{r}}} =  - \dfrac{{a_{\rm{r}}^2}}{{12}} - {c_{\rm{r}}},\\
    {Q_{\rm{r}}} = \dfrac{1}{3}{a_{\rm{r}}}{c_{\rm{r}}} - \dfrac{{a_{\rm{r}}^3}}{{108}} - \dfrac{{b_{\rm{r}}^2}}{8}.
    \end{array}
    \end{equation}
Thus, the exact precessional angular frequency modified by inertia is given by
    \begin{equation} \label{eq:wp}
    \begin{split}
	&\omega _{\rm{p}}^{} =  \frac{{\sqrt {{a_{\rm{r}}} + 2{y_{\rm{r}}}} }}{2} \\
	&- \frac{1}{2}\sqrt { - 3{a_{\rm{r}}} - 2{y_{\rm{r}}} - \frac{{2{b_{\rm{r}}}}}{{\sqrt {{a_{\rm{r}}} + 2{y_{\rm{r}}}} }}},
	\end{split}
	\end{equation}
The exact nutational angular frequency can be written as
    \begin{equation} \label{eq:wn}
    \begin{split}
	&\omega _{\rm{n}}^{} = \frac{{\sqrt {{a_{\rm{r}}} + 2{y_{\rm{r}}}} }}{2} \\
	&+ \frac{1}{2}\sqrt { - 3{a_{\rm{r}}} - 2{y_{\rm{r}}} - \frac{{2{b_{\rm{r}}}}}{{\sqrt {{a_{\rm{r}}} + 2{y_{\rm{r}}}} }}}.
	\end{split}
	\end{equation}

%\begin{figure*}[t]
%    \centering
%    \includegraphics[width=1\textwidth]{220228 FigError.png}
%    \caption{The errors of approximate solutions of Eqs. (\ref{eq:wp}) and (\ref{eq:wn}) for the configurations shown in Fig. \ref{fig:FreqH}. (a)-(d)~The approximation ”a” of precession frequency. (e)-(h)~The approximation ”b” of precession frequency. (i)-(l)~The approximation ”a” of nutation frequency. (m)-(p)~The approximation ”b” of precession frequency.}
%    \label{fig:Error}
%\end{figure*}

Next, we write a few approximations allowing one to elucidate the physics behind Eqs. (\ref{eq:wp})-(\ref{eq:wp}). The approximation "a" of resonance frequencies is derived by taking into account the real part of the quartic equation~(\ref{eq:SBC_appendix}), which transforms this equation into a bi-quadratic one. Thus, the approximation "a" of precessional frequency reads
    \begin{equation} \label{eq:w_approx_a_prec_appendx}
	\omega _{\rm{p}}^{\left( {\rm{a}} \right)} =  {\left( { p - \sqrt {{{p}^2} - q} } \right)^{1/2}}.
	\end{equation}
The approximate nutational frequency is given by
    \begin{equation} \label{eq:w_approx_a_nut_appendx}
	\omega _{\rm{n}}^{\left( {\rm{a}} \right)} = {\left( { p + \sqrt {{{p}^2} - q} } \right)^{1/2}},
	\end{equation}
where
    \begin{equation} \label{eq:pq_appendx}
	\begin{array}{l}
    p =  \dfrac{1}{{{2 \eta ^2}}} + \dfrac{\Delta\omega_{\rm{SB}}}{2 \alpha\eta},\\
    q = \dfrac{\omega_{\rm{SB}}^2}{\eta^2},
    \end{array}
    \end{equation}
This approximation introduces an additional error, which does not exceed 0.5\% for the aligned modes for the parameters employed in the main part of the paper. 
%The relative error diverges (Fig.\,\ref{fig:Error}(c,g)) only for magnetic fields with precessional frequencies approaching zero (division by zero). 
We thus find that the solution 'a' can be considered sufficiently accurate in the context of this work.

Expressions (\ref{eq:w_approx_a_prec_appendx}) and (\ref{eq:w_approx_a_nut_appendx}) can be further simplified by employing Taylor series expansion and assumption that $\Delta\omega_{\rm{SB}}\eta  \ll 1$. The approximation 'b' of precessional frequency is
    \begin{equation} \label{eq:w_approx_b_prec_appendx}
    \begin{split}
% 	&\omega_{\rm{p}}^{\left( {\rm{b}} \right)} = \left[ {\frac{{{{\left| \gamma  \right|}^2}\left( {1 + {\alpha ^2}} \right)}}{{M_0^2{{\sin }^2}{\theta _0}}}\left( {{\partial _{\theta \theta }}F{\partial _{\phi \phi }}F - {{\left( {{\partial _{\theta \phi }}F} \right)}^2}} \right)
% 	\\
% 	&- \eta \frac{{{{\left| \gamma  \right|}^3}\left( {1 + {\alpha ^2}} \right)}}{{M_0^3{{\sin }^2}{\theta _0}}}\left( {{\partial _{\theta \theta }}F{\partial _{\phi \phi }}F - {{\left( {{\partial _{\theta \phi }}F} \right)}^2}} \right) \times \\
% 	&\left( {{\partial _{\theta \theta }}F + \frac{{{\partial _{\varphi \varphi }}F}}{{{{\sin }^2}{\theta _0}}}} \right)} \right]^{1/2}.
	{\omega_{\rm{p}}^{(\rm{b})}} = \omega_{\rm{SB}} \sqrt{1-\eta\frac{\Delta\omega_{\rm{SB}}}{\alpha}}.
	\end{split}
	\end{equation}
From Eq.~(\ref{eq:w_approx_b_prec_appendx}), one can see that this expression of precessional frequency modified by inertia converges to the conventional expression of FMR at $\eta  = 0$. The approximation "b" of the nutational frequency reads
    \begin{equation} \label{eq:w_approx_b_nut_appendx}
% 	\omega _{\rm{n}}^{\left( {\rm{b}} \right)} = \frac{1}{\eta } + \frac{{\left| \gamma  \right|\left( {{\partial _{\phi \phi }}F + \left( {\partial _{\theta \theta }}F \right) {{\sin }^2}\theta } \right)}}{{2{M_0}{{\sin }^2}{\theta _0}}}.
	{\omega} _{\rm{n}}^{(\rm{b})} = \frac{1}{\eta } + \frac{\Delta\omega_{\rm{SB}}}{2\alpha}.
	\end{equation}
%Note that in the high-anisotropy ferromagnets, the second term in Eq. (\ref{eq:w_approx_b_nut_appendx}). can cause significant dependence of nutation frequency on magnetic field. 
The series expansion leads to a further error of about 6\% for the parameters used for numerical calculations presented in the main part of the paper. 

\section{FREE ENERGY DENSITY}
\label{suppl3}
\begin{figure}[t]
    \centering
    \includegraphics[width=0.5\textwidth]{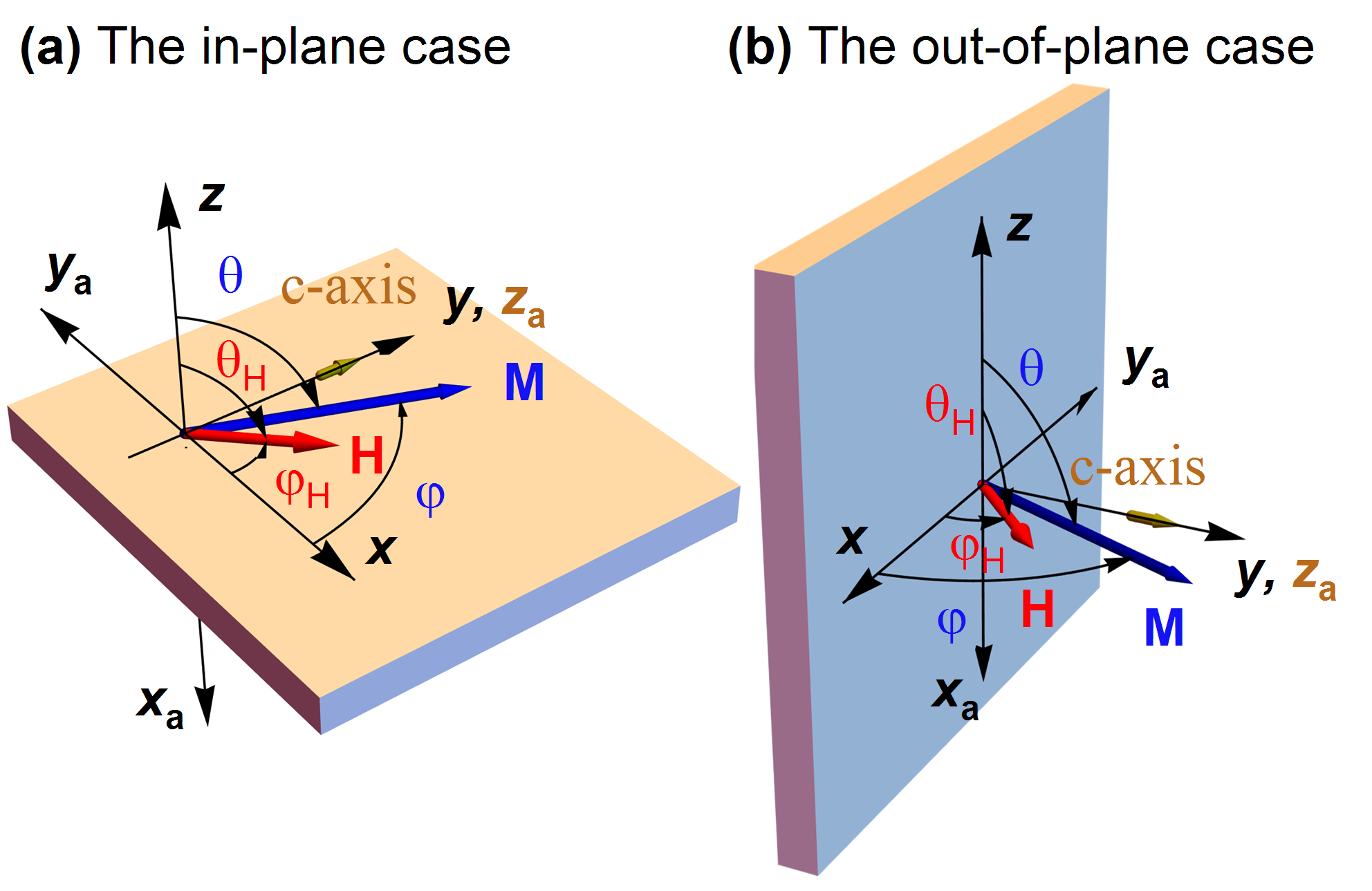}
    \caption{The orientation of coordinate frames in the case of magnetic field applied in-plane (a) and out-of-plane (b).
    }
    \label{fig:analytic_angles}
\end{figure}

We consider two geometrical configurations are of interest for the study of resonances in magnetic materials. In the first configuration, a magnetic field rotates tangentially through the plane of the film surface the \emph{x0y} plane (Fig.\,\ref{fig:analytic_angles}(a)). In the second configuration, the magnetic field goes from the tangential direction to the normal direction in the \emph{x0y} plane (Fig.\,\ref{fig:analytic_angles}(b)). In the geometries selected here, the films are located differently relative to the axes, thus the aforementioned planes do not coincide. Such choice of the axes allows one to avoid the division by zero in the out-of-plane applied field configuration (Fig.\,\ref{fig:analytic_angles}(b)). Otherwise, if one directs the magnetic field along the normal to the film in the configuration of axes shown in Fig.~\ref{fig:analytic_angles}(a), one obtains singularity $\left( {{\theta _0} = 0} \right)$ in Eqs.\,(\ref{eq:SBC}) and (\ref{eq:SB}). Note that an alternative approach was derived in the past by Baselgia et al. \cite{Baselgia_Warden1988}.
\begin{figure}[t]
    \centering
    \includegraphics[width=0.5\textwidth]{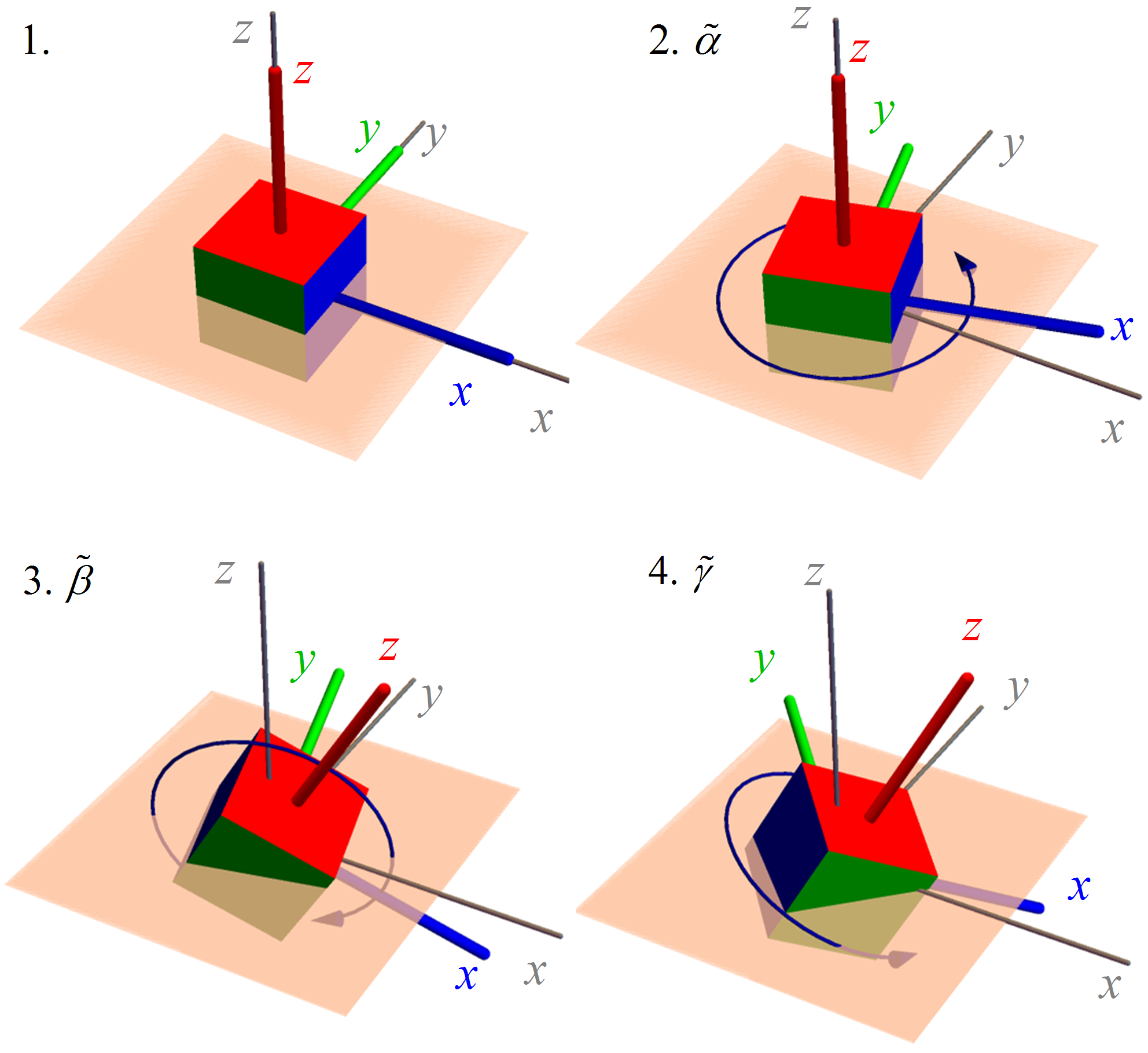}
    \caption{The sequence of z-y-z rotations to the $\tilde \alpha ,$ $\tilde \beta $ and $\tilde \gamma $ angles respectively.
    }
    \label{fig:Eule_rotation}
\end{figure}

In order to write expressions of the energy contributions, let us define coordinate frames. In the general case, the axes of magnetocrystalline anisotropy (cubic, uniaxial, etc) may not coincide with the demagnetization axes, therefore, one needs to make a transition from one axis to another to calculate the energy. 
We indicate angles of magnetization vector as $\theta $ and $\varphi $ respectively to Cartesian coordinate system \emph{xyz}, defining the demagnetizing energy. The polar and azimuthal angles of the vector of the magnetic field are denoted by ${\theta _{\rm{H}}}$ and ${\varphi _{\rm{H}}}$ with respect to the same axes. The axes specifying the energy of magnetocrystalline anisotropy are indicated by ${x_{\rm{a}}}{y_{\rm{a}}}{z_{\rm{a}}}$. For instance, we focus on uniaxial anisotropy in the rotated coordinate system such that the c-axis $\left( {{z_a}} \right)$  is aligned with the y-axis. Thus, the free energy density is given by
    \begin{equation} \label{eq:F_all}
	F = {F_{\rm{Z}}} + {F_{{\rm{dm}}}} + {F_{\rm{a}}},
	\end{equation}
where ${F_{\rm{Z}}}$ is the Zeeman energy density, ${F_{{\rm{dm}}}}$ is the demagnetizing energy density and ${F_{\rm{a}}}$ is related to magnetocrystalline anisotropy. Using representation of vectors in a spherical coordinate system, one writes the Zeeman energy as
	\begin{equation} \label{eq:F_Zeeman}
	\begin{split}
	F_{\rm{Z}} &=  - \mu_0 \bf{MH} \\
	&=  - \mu_0 M_0 H \left( \right. \sin \theta \sin \theta_{\rm{H}} \cos \left( \varphi  - \varphi_{\rm{H}} \right) \\
	& + \cos \theta \cos \theta_{\rm{H}} \left. \right)
	\end{split}
	\end{equation}
and the demagnetizing energy as
    \begin{equation} \label{eq:F_dm}
    \begin{split}
	F_{\rm{dm}} = \frac{1}{2} \mu_0 M_0^2 \left( \right. N_{\rm{x}} \sin^2 \theta \cos ^2\varphi  
	& + N_{\rm{y}} \sin^2\theta \sin^2 \varphi  \\
	& + N_{\rm{z}} \cos^2\theta \left. \right),
	\end{split}
	\end{equation}
where ${N_{\rm{x}}},$ ${N_{\rm{y}}},$ ${N_{\rm{z}}}$ are demagnetizing factors. The magnetocrystalline anisotropy energy of a ferromagnet with cubic symmetry is given by
    \begin{equation} \label{eq:F_cubic}
	\begin{split}
    {F_{{\rm{cub}}}} &=  K_{{\rm{cub1}}}^{}\left( {\alpha _1^2\alpha _2^2 + \alpha _2^2\alpha _3^2 + \alpha _1^2\alpha _3^2} \right) \\ 
    &+ K_{{\rm{cub2}}}^{}\alpha _1^2\alpha _2^2\alpha _3^2 + ... \\[4pt]
    &= K_{{\rm{cub1}}}^{}{\cos ^2}{\varphi _{\rm{a}}}{\sin ^2}{\theta _{\rm{a}}}\left[ {{{\cos }^2}{\varphi _{\rm{a}}}}  { + \left( {1 + {{\sin }^2}{\theta _{\rm{a}}}} \right){{\sin }^2}{\varphi _{\rm{a}}}} \right] \\
    &+ K_{{\rm{cub2}}}^{}{\sin ^4}{\theta _{\rm{a}}}{\sin ^2}{\varphi _{\rm{a}}}{\cos ^4}{\varphi _{\rm{a}}} + ...,
 	\end{split}
	\end{equation}
where ${\theta _{\rm{a}}}$ and ${\varphi _{\rm{a}}}$ are polar and azimuthal angles of the magnetization vector in the ${x_{\rm{a}}}{y_{\rm{a}}}{z_{\rm{a}}}$ frame, and ${\alpha _1},$ ${\alpha _2}$ and ${\alpha _3}$ are directional cosines with respect to the same frame. Finally, the uniaxial anisotropy energy can be written as
    \begin{equation} \label{eq:F_uni}
	\begin{split}
	{F_{{\rm{uni}}}} &= K_{{\rm{u1}}}^{}{\sin ^2}{\theta _{\rm{a}}} + K_{{\rm{u2}}}^{}{\sin ^4}{\theta _{\rm{a}}} + K_{{\rm{u3}}}^{}{\sin ^6}{\theta _{\rm{a}}} \\
	&+ K_{{\rm{u4}}}^{}{\sin ^6}{\theta _{\rm{a}}}\cos 6{\varphi _{\rm{a}}} + ...,
	\end{split}
	\end{equation}
where constants of anisotropy are denoted with ${K_i}$.

The magnetization vector can be specified in two equivalent ways
    \begin{equation} \label{eq:M}
	\begin{array}{l}
	{\bf{M}} = M\left[ \begin{array}{l}
    {\alpha _1}\\
    {\alpha _2}\\
    {\alpha _3}
    \end{array} \right] = M\left[ \begin{array}{l}
    \sin {\theta _{\rm{a}}}\cos {\varphi _{\rm{a}}}\\
    \sin {\theta _{\rm{a}}}\sin {\varphi _{\rm{a}}}\\
    \cos {\theta _{\rm{a}}}
    \end{array} \right],
    \end{array}
	\end{equation}
therefore, one can write
    \begin{equation} \label{eq:thetaa_varphia}
	\begin{array}{l}
    {\theta _{\rm{a}}} = \arccos {\alpha _3},\\
    {\varphi _{\rm{a}}} = \arctan \dfrac{{{\alpha _2}}}{{{\alpha _1}}}.
    \end{array}
    \end{equation}
On the other hand, one can match the vector components in the ${x_{\rm{a}}}{y_{\rm{a}}}{z_{\rm{a}}}$ frame with the $xyz$ frame using the Euler rotation matrix in the form
    \begin{equation} \label{eq:alpha_Euler}
	\left[ \begin{array}{l}
    {\alpha _1}\\
    {\alpha _2}\\
    {\alpha _3}
    \end{array} \right] = {E_{\rm{u}}}{\left( {\tilde \alpha ,\tilde \beta ,\tilde \gamma } \right)^T}\left[ \begin{array}{l}
    \sin \theta \cos \varphi \\
    \sin \theta \sin \varphi \\
    \cos \theta 
    \end{array} \right],
    \end{equation}

    \begin{equation} \label{eq:Euler}
    \begin{split}
	&{E_{\rm{u}}}\left( {\tilde \alpha ,\tilde \beta ,\tilde \gamma } \right) \\
	& = \left[ {\begin{array}{*{20}{c}}
    {{c_{\tilde \alpha }}{c_{\tilde \beta }}{c_{\tilde \gamma }} - {s_{\tilde \alpha }}{s_{\tilde \gamma }}}&{ - {c_{\tilde \gamma }}{s_{\tilde \alpha }} - {c_{\tilde \alpha }}{c_{\tilde \beta }}{s_{\tilde \gamma }}}&{{c_{\tilde \alpha }}{s_{\tilde \beta }}}\\
    {{c_{\tilde \beta }}{c_{\tilde \gamma }}{s_{\tilde \alpha }} + {c_{\tilde \alpha }}{s_{\tilde \gamma }}}&{{c_{\tilde \alpha }}{c_{\tilde \gamma }} - {c_{\tilde \beta }}{s_{\tilde \alpha }}{s_{\tilde \gamma }}}&{{s_{\tilde \alpha }}{s_{\tilde \beta }}}\\
    { - {c_{\tilde \gamma }}{s_{\tilde \beta }}}&{{s_{\tilde \beta }}{s_{\tilde \gamma }}}&{{c_{\tilde \beta }}}
    \end{array}} \right],
    \end{split}
    \end{equation}
thereby one rotates the $xyz$ axes to the ${x_{\rm{a}}}{y_{\rm{a}}}{z_{\rm{a}}}$ axes. Note that the coordinate system rotates, not the vector, therefore the Euler matrix is transposed. Here we introduce the short notations for trigonometric functions ${c_{\tilde \alpha }} = \cos \tilde \alpha ,$ ${s_{\tilde \alpha }} = \sin \tilde \alpha $ and so on. The given Euler rotation matrix describes a sequence of rotations to the angles $\tilde \alpha ,$ $\tilde \beta $ and $\tilde \gamma $ around the z, y and z local axes (Fig. \ref{fig:Eule_rotation}). Thus, one can express directional cosines ${\alpha _1},$ ${\alpha _2}$ and ${\alpha _3}$ through the predetermined rotation angles $\tilde \alpha ,$ $\tilde \beta ,$ $\tilde \gamma $ and angles of magnetization $\theta $ and $\varphi $ in the $xyz$ frame; then one can use formulas (\ref{eq:thetaa_varphia}) and the corresponding expression of energy density to calculate anisotropy energy in the the $xyz$ coordinate system.

For example, we focus on ferromagnets with uniaxial symmetry and cases shown in Fig. \ref{fig:analytic_angles}, then the consistency between the magnetization angles in $xyz$ and ${x_{\rm{a}}}{y_{\rm{a}}}{z_{\rm{a}}}$ frames is given by
    \begin{equation} \label{eq:Euler_example1}
	\begin{split}
    {E_{\rm{u}}}{\left( {\frac{\pi }{2},\frac{\pi }{2},0} \right)^T} = \left[ {\begin{array}{*{20}{c}}
    0&0&{ - 1}\\
    { - 1}&0&0\\
    0&1&0
    \end{array}} \right],
    \end{split}
    \end{equation}
    \begin{equation} \label{eq:Euler_example2}
    \left[ \begin{array}{l}
    {\alpha _1}\\
    {\alpha _2}\\
    {\alpha _3}
    \end{array} \right] = \left[ \begin{array}{l}
    - \cos \theta \\
    - \sin \theta \cos \varphi \\
    \sin \theta \sin \varphi 
    \end{array} \right],
    \end{equation}
    \begin{equation} \label{eq:Euler_example3}
    \begin{split}
    {\theta _{\rm{a}}} = \arccos \left( {\sin \theta \sin \varphi } \right),\\
    {\varphi _{\rm{a}}} = \arctan \left( {\tan \theta \cos \varphi } \right).
    \end{split}
    \end{equation}
The energy density is defined as
    \begin{equation} \label{eq:F_example}
    \begin{split}
	F & =  - \mu_0 M_0 H \left( \right. \sin \theta \sin \theta_H \cos \left( \varphi  - \varphi_H \right) \\
	& + \cos \theta \cos \theta_H \left. \right) \\
	& + K_{\rm{u1}} \left( 1 - \sin^2 \theta \sin^2\varphi \right) + F_{\rm{dm}},
	\end{split}
    \end{equation}
where we neglect the high-order anisotropy terms. For the in-plane configuration (Fig. \ref{fig:analytic_angles}(a)), one writes
    \begin{equation} \label{eq:F_IP}
	{F_{{\rm{dm}}}} = \frac{{{\mu _0}M_0^2}}{2}{\cos ^2}\theta,
	\end{equation}
whereas for the out-of-plane case shown in Fig. \ref{fig:analytic_angles}(b), the demagnetizing energy is given by
    \begin{equation} \label{eq:F_OOP}
	{F_{{\rm{dm}}}} = \frac{{{\mu _0}M_0^2}}{2}{\sin ^2}\theta {\sin ^2}\varphi.
	\end{equation}
The presented expressions of energy density allow one to avoid the division by zero in the out-of-plane magnetization configuration $\left( {{\theta _0} = 0} \right)$. One can calculate the second derivatives of the energy density and substitute the results in Eq. (\ref{eq:SBC}) to find the FMR frequency modified by inertia or the nutation frequency.

\section{EQUILIBRIUM ANGLES OF MAGNETIZATION}
\label{suppl4}

\begin{figure*}[htp]
    \centering
    \includegraphics{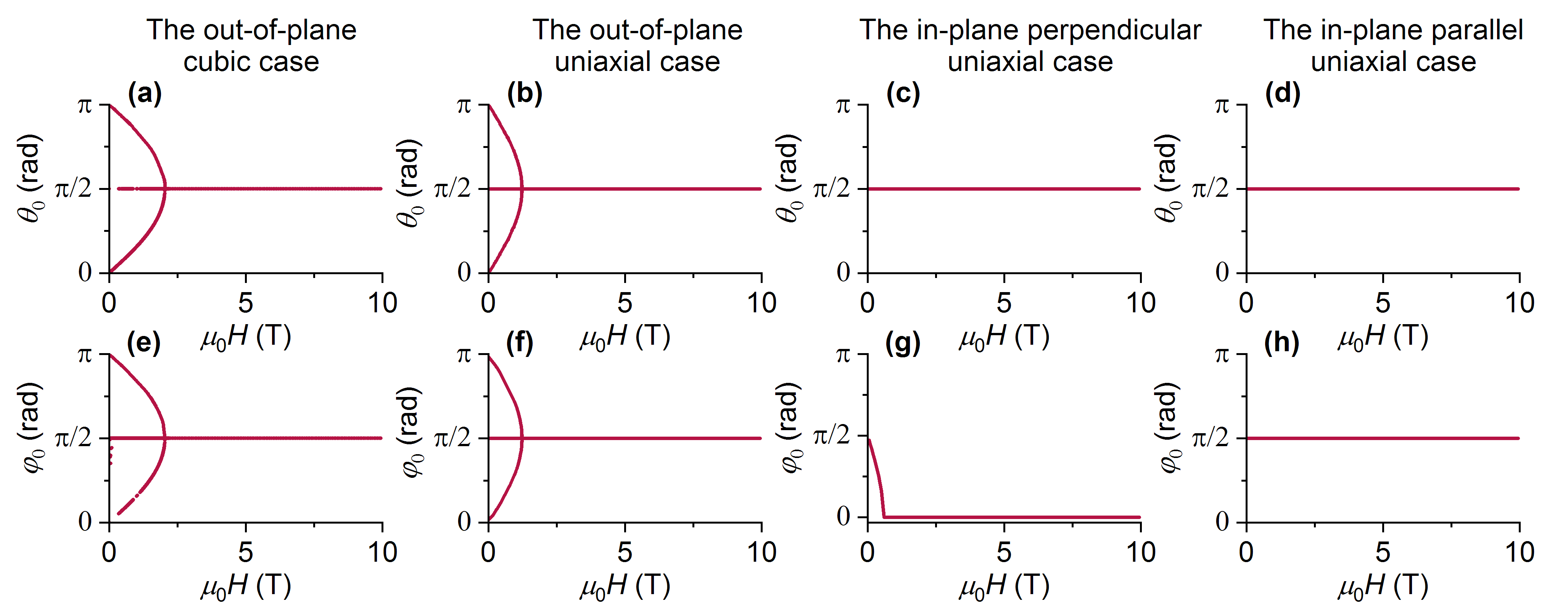}
    \caption{Equilibrium angles of magnetization. (a-d) polar angles. (e-h) azimuthal angles. }
    \label{fig:Angles}
\end{figure*}

Based on the presented approach, we find equilibrium angles of magnetization for all the investigated configurations and the results are plotted in Fig.\,\ref{fig:Angles}. Note that the angles for the in-plane cases are calculated for the geometry shown in Fig. \ref{fig:analytic_angles}(a), while the angles for out-of-plane cases are given in other axes (Fig.\,\ref{fig:analytic_angles}(b)).
\bibliography{references}

\end{document}